\documentclass[preprint,prX,floatfix]{revtex4-1}
\usepackage{bm,amsmath,amssymb,mathrsfs}
\usepackage{cancel,epsfig,graphicx,wasysym}
\usepackage{tikz}
\usetikzlibrary{arrows,snakes,shapes}
\begin{document}
\title{Parity nonconserving cold neutron-parahydrogen
interactions}
\author{T.~M.~Partanen}
\email{tero.partanen@helsinki.fi}
\affiliation{ Department of Physical Sciences,P.~O.
~Box~64, FIN-00014 University of Helsinki, Finland}
%
%
\begin{abstract}
Three pion dominated observables of the parity
nonconserving interactions between the cold neutrons
and parahydrogen are calculated. The transversely
polarized neutron spin rotation, unpolarized neutron
longitudinal polarization, and photon-asymmetry of the
radiative polarized neutron capture are considered.   
For the numerical evaluation of the observables, the
strong interactions are taken into account by the
Reid93 potential and the parity nonconserving 
interactions by the DDH model along with the two-pion
exchange. 
\end{abstract}
\maketitle
\section{Introduction}
The knowledge of the strangeness conserving hadronic
weak interaction relies completely on the parity
nonconserving (PNC) observables. The PNC two-nucleon
($NN$) interactions provide a possible access to this
least understood sector of the Standard Model.
Experimentally such subtle particle spin control based
PNC measurements are feasible but highly demanding. 
On the theoretical side, the challenge lies largely in
the poorly known coupling values which parametrize the
strength of the  minuscule-sized PNC signal.

The PNC $NN$ interaction is compelled to change
either the spin or isospin of the system due to the
Pauli exclusion principle and consequently the 
potential of the interaction is composed of various
spin-isospin operators weighted by coupling constants. 
There are at least three considerable candidates for
the PNC $NN$ potentials. The new model-independent
effective field theory (EFT) approach offers two
alternative choices for these potentials, namely the
pionless and pionful \cite{zhuetal,rammus}. 
The pionless one comprises only the short-range contact
interaction whereas the pionful also the long- and
medium-range interactions mediated respectively by the
one- and two-pion exchanges. The third potential is the
most conventional DDH meson-exchange model \cite{ddh}
which takes into account the long- and short-range
effects in terms of the single $\pi^{\pm}, \rho,$ and
$\omega$ exchanges but not the two-pion exchange
contributions which are supposedly important in the
medium-range.
In spite of the different approaches, the operators
appearing in the potentials are essentially the same,
except that the ranges of the force in them varies.
Another similarity with the potentials is that they
are all parametrized in terms of about half a dozen
ill-known couplings. 
In EFT these so-called low energy constants are 
expected to be extracted from the experimental data of
a series of prospective high-precision measurements.
As a downside, the EFT potentials cannot be used in the
evaluation of observables yet. Therfore, for over thirty
years up until today, the theoretical predictions have
largely rest on the DDH model and their recommended
"best" values for the weak couplings.
The PNC one-pion exchange potentials of the pionful EFT
and DDH model coincide and are proportional to the weak
$NN\pi$ coupling $h^{(1)}_\pi$. 
The strength of the PNC two-pion exchange is also
dependent on the same coupling. Besides the DDH, there
are various calculations 
\cite{fcdh,desp80,dubo86,iqbal89,kais89,henley98,lobov02,
leehj} for the the $h_\pi^{(1)}$ (ranging between $0$ and
$3.4\times10^{-7}$) indicating a smaller value than what
is the DDH "best" recommendation. The ongoing NPDGamma
experiment \cite{npdgamma} is hoped to reduce the
vagueness related to the $h_\pi^{(1)}$ by measuring with
high accuracy the $\gamma$-asymmetry of the
$\vec{n}p\rightarrow\gamma d$ at threshold known to be
nearly a $100\%$ pion exchange dominated.

Up till recently, the effect of the two-pion exchange
has been considered small and neglected from the 
analyses of the PNC observables.
However, recent calculations of the PNC longitudinal
analyzing power $\bar{A}_L$ in the $\vec{p}p$ elastic
scattering show the importance of the two-pion exchange
\cite{pp1,pp2}. The PNC $pp$ reaction offers an
auspicious opportunity to study the two-pion exchange
contribution for a couple of reasons. Firstly, the single
pion-exchange does not appear according to Barton's
theorem \cite{barton}, from which it follows that the 
two-pion exchange represents the longest-ranged and
probably dominated contribution. Secondly, there
exist three high-precision measurements of the
$\bar{A}_L$ at different energies (Bonn at 13.6 MeV, 
PSI at 45 MeV, and TRIUMF at 221.3 MeV) which can be
compared to the theoretical predictions. Consequently,
when it comes to the one-pion exchange dominant PNC $np$
interactions, the effect of the two-pion exchange
should also be counted in.

When slow neutrons collide with hydrogen molecules, 
they either elastically scatter off or get absorbed in
the protons resulting in deuterons and photons. 
This work presents calculations of the three different
pion sensitive PNC observables arising from the cold
neutron interaction with parahydrogen. 
Two are due to the PNC elastic scattering enabling
the spin rotation $\frac{d}{dz}\phi$ and polarization
$\frac{d}{dz}P$ of the neutrons and the third one is 
the $\gamma$-asymmetry $\mathcal{A}_\gamma$ in the 
radiative PNC capture of polarized neutrons.
The spin rotation and polarization of the neutron in 
the PNC $np$ scattering were first discussed in Refs.
\cite{michel,stods} and the wavefunction based
calculations have later been performed in Refs.
\cite{avis2,schia,liutimm,liu07}.
There has also been experimental interest in measuring
the PNC neutron spin rotation $\frac{d}{dz}\phi$ in
a liquid parahydrogen target at the Neutron Spallation
Source (SNS), see Ref. \cite{markoff}.
The radiative PNC reaction $\vec{n}p\rightarrow d\gamma$
is also discussed in multiple papers, of which 
Refs \cite{tadic,danilov,desp,lassey,morioka,kaplan,
despnp,savage,hyunpark,schiacar,schia,haid,hyunncap,
liu07,hyandes,ando,part} present numerical predictions.  

In the present work the Reid93 potential \cite{reid93}
is chosen to take care of the strong $np$ interactions.
The long- and medium-range parts of the weak interaction
are respectively due to the one- and two- pion exchanges.
The contributions of the two-pion exchanges are taken
into account in the observables by using the two 
separate PNC NN two-pion exchange potentials taken from
Refs. \cite{k2pi,d2pi}. The former of these potentials
counts in the $\Delta(1232)$ isobar effects while the
latter one does not. The neutron spin rotation
$\frac{d}{dz}\phi$ and polarization $\frac{d}{dz}P$
are given in terms of the reduced matrix elements of
the spin-space operators, which are the basic 
building blocks of the PNC potentials. The radial 
Yukawa functions of the operators are used in
unregularized form and parameterized in three different
ways by the meson masses $\pi$, $\rho$, and $\omega$.
The calculations are performed in the distorted-wave
Born approximation (DWBA), in which the phenomenological
strong interaction wavefunctions sandwich the PNC
operator. In the calculation of the $\gamma$-asymmetry
$\mathcal{A}_\gamma$ the pion-exchange current effects
are included in the form of two-body dipole operators. 
The required bound and continuum radial wavefunctions 
together with their tiny parity admixed components are
obtained from the exact solution of the coupled 
Schr\"{o}dinger equation. 

The remainder of the paper is organized as follows.
Section \ref{form} gives the basic formalism of the 
cold neutron interaction with parahydrogen, Sec.
\ref{res} presents the results of the calculated
observables, and Sec. \ref{summar} summarizes the work. 
\section{\label{form}Formalism}
Hydrogen exists in nature in molecular form, composed 
of two protons bound by two electrons. The hydrogen
molecule ($H_2$) comes in two species called 
parahydrogen ($H_2^{\rm p}$) and orthohydrogen 
($H_2^{\rm o}$) with proton spins aligned antiparallel
and parallel respectively. 
The de Broglie wavelength of a neutron at energies of
a few meV is much greater than the internuclear 
separation $R_0=0.75~\mathring{A}$ between the protons
in the hydrogen molecule. Consequently the neutrons,
if not captured by protons, scatter coherently off the
two protons in the molecules. Parahydrogen molecule is
spinless, since its proton spins couple to zero and,
therefore, it cannot depolarize a polarized neutron 
when they scatter elastically. 
The protonic wavefunction of the hydrogen molecule must
be antisymmetric from which follows that in the ground
states of the para- and orthohydrogen molecules, the
rotational energies are respectively zero and 
$I^{-1}=14.7$ meV, where $I=\mu R_0^2$ is the moment
of inertia and $\mu$ the reduced mass of two protons. 
This energy determines the upper limit of the neutron
center of mass (C.M.) energy in order not to get
depolarized by the conversion of the para- to
orthohydrogen molecule.

The low energy neutron-parahydrogen ($nH_2^{\rm p}$)
interaction is a three-body problem that can crudely be
simplified to a two-body neutron-proton ($np$) interaction
problem. The $np$ continuum wavefunctions, in which the
z-axis is taken along the direction of $\bm{k}$, are of
the form
\begin{align}\label{pnwf}
\langle\bm{r}|k\hat{\bm{z}};\mathcal{Q}
m_n\rangle^{(\pm)}=~&
\frac{\sqrt{4\pi}}{kr}
i^{L}\sqrt{2L+1}\sum_{L'm_p}
\langle{\textstyle\frac{1}{2}}m_n
{\textstyle\frac{1}{2}}m_p\vert SM_S\rangle
\nonumber\\
\times&\langle L0SM_S\vert JM_S\rangle
\mathcal{U}_{\mathcal{Q}L'}^{(\pm)}(k,r)
\mathscr{Y}^{L'S}_{JM_S}(\hat{\bm{r}})
(-)^{T+1}\vert T0\rangle,
\end{align}
where the superscripts $(\pm)$ refer to the incoming
$(-)$ and outgoing $(+)$ wave boundary conditions,
$\mathscr{Y}^{L'S}_{JM_S}(\hat{\bm{r}})$ are the
eigenfunctions of the coupled angular momentum, and
the quantum numbers $LSJT$ are abbreviated to
$\mathcal{Q}$. The quantum numbers $STJ$ do not change
under strong interaction. Because of the
antisymmetricity requirement of the wavefunction, the
isospin $T$ may as well be considered uniquely defined
by the $LS$, and so, on occasion, the $\mathcal{Q}$ is
also designated for convenience with the spectroscopic
notation $\mathcal{Q}={}^{2S+1}L_J$. 
An adequate deuteron wavefunction, in the 
$\gamma$-asymmetry $\mathcal{A}_\gamma$, is composed of
three partial waves, which are the usual tensor coupled
${}^3S_1-{}^3D_1$ and tiny parity admixed ${}^3P_1$. 
The $np$ bound state wavefunction is given as
\begin{equation}
\langle\bm{r}|M_d\rangle=
\sum_{L_d}\frac{\mathcal{D}_{L_d}(r)}{r}
\mathscr{Y}^{L_d1}_{1M_d}(\hat{\bm{r}})
\vert T_d0\rangle,
\end{equation}
with the normalization 
$\int_0^\infty dr\sum_i^3|\mathcal{D}_i|^2=1$ and
energy eigenvalue of $-2.2246$ MeV.
\subsection{Neutron scattering}
Since the interest here is in the coherent $nH_2^p$
scattering, one must carefully take into account the
relative motion between the neutron and system of the
chemically bound protons.
When a low energy neutron comes across the molecules in
the medium, it interacts collectively with a number of
them. As a result, the scattered waves originating from
the molecules, interfere with the through passing neutron
and change its momentum.
Applying the Lippmann-Schwinger equation for multiple
point-like scatterers each located at $\bm{r}_j$,
the wave of a slow neutron after travelling through the 
target in the z-direction then becomes 
\begin{equation}\label{nhwf}
e^{iq'z}\approx e^{iqz}
+\tilde{f}(q,0)
\sum_j\frac{e^{iq|\bm{r}-\bm{r}_j|}}
{|\bm{r}-\bm{r}_j|}e^{iqz_j},
\end{equation}
where $\tilde{f}(q,\theta=0)$ denotes the forward
$nH_2^p$ scattering amplitude and $q$ is the relative
momentum of the neutron and molecule. 
The sum of the spherical waves from the scatterers in 
Eq. \eqref{nhwf} may be written in the form of an 
integral over a smooth distribution of scattering 
centers in a cylindrical shaped target of infinite
radius. For the neutrons travelling along the axis of
the target, it then follows that the right hand side
of Eq. \eqref{nhwf} becomes $e^{i(qz-\varphi)}$, with
\begin{equation}\label{rota}
\varphi(q,z)=
-\frac{2\pi\mathcal{N}z}{q}
\tilde{f}(q,0),
\end{equation}
where $\mathcal{N}$ is the particle density of the
medium.
The neutron thus gains the $\varphi=(q-q')z$
amount of phase when propagating through a medium of
length $z$. Equation \eqref{rota} is related to the
index of refraction $n=q'/q=1-\varphi/qz$ in neutron
optics.

By the initial choice of the transversely polarized
spin in the positive x-direction 
$\langle\sigma_x\rangle=+1$, 
the neutron spin wavefunction 
$|x+\rangle=(|+\rangle+|-\rangle)/\sqrt{2}$
contains equal amount of $\pm$ helicities in the
direction of its propagation along the z-axis.
The PNC interaction favors one helicity state slightly
more than the other and thus depending on this state,
the neutrons scatter a bit differently. The neutron
wavefunction accumulates the $\varphi_{m_n}$ amount
phase (where $m_n$ is the spin polarization of the
incident neutron) labeled individually for each two
states when passing through the target. 
It follows straightforwardly from the expectation
value of the spin $\langle\bm{\sigma\rangle}$, that
the neutron spin rotates in the xy-plane around the
z-axis if the real value of the subtraction between
the helicity states of Eq. \eqref{rota} is non-zero 
\begin{equation}\label{rotang}
\phi(q,z)=-\frac{2\pi\mathcal{N}z}{q}
{\rm Re}\Bigl(\tilde{f}_{+\frac{1}{2}}(q,0)
-\tilde{f}_{-\frac{1}{2}}(q,0)\Bigr).
\end{equation}
In the case of the unpolarized neutron beam, the 
neutrons gain some amount of longitudinal polarization
due to the parity nonconservation when propagating 
through a medium. The incident beam intensity loss is
given by
$dI_\pm(q,z)=-\mathcal{N}\sigma_\pm(q)I_\pm(q,z)dz$
from which, with the help of the optical theorem, the
fractional polarization follows as the difference 
between the $I_+(q,z)$ and $I_-(q,z)$ divided by their
sum
\begin{align}\label{pola}
P(q,z)&
\approx-\frac{2\pi\mathcal{N}z}{q}
{\rm Im}\Bigl(\tilde{f}_{+\frac{1}{2}}
(q,0)-\tilde{f}_{-\frac{1}{2}}(q,0)\Bigr).
\end{align}  

The $nH_2^p$ scattering amplitude may be written as
$\tilde{f}=-(\tilde{\mu}/\mu)\tilde{a}$ in
which the $\mu\approx M/2$ and 
$\tilde{\mu}\approx 2M/3$ are respectively the 
neutron-proton and neutron-molecule reduced masses 
with the average nucleon mass $M=939$ MeV and
$\tilde{a}=(a_s+3a_t)/2$ (see {\it e.g.} 
Ref. \cite{blatt}) is the coherent scattering length 
expressed in terms of the $np$ scattering lengths for
the singlet ${}^1S_0$ and triplet ${}^3S_1$ channels.
That is $\tilde{a}=-f/2$, where $f$ is the $np$
scattering amplitude.
The $nH_2^p$ and $np$ scattering amplitudes and 
momenta are related as $\tilde{f}=(\tilde{\mu}/2\mu)f$
and $q=(\tilde{\mu}/\mu)k$, where $k$ is the relative
momentum of the neutron and proton.
The relevant PNC part of the forward $np$ scattering
amplitude $f_{m_n}(k,\theta=0)$ in distorted wave Born
approximation (DWBA) is given by
\begin{align}
f_{m_n}(k,0)=
-\frac{\mu}{2\pi}
{}^{(-)}\langle k\hat{\bm{z}};m_n|
\hat{V}^{{\rm PNC}}|
k\hat{\bm{z}};m_n\rangle^{(+)},
\end{align}
where the matrix elements are Hermitian. 
The $S\leftrightarrow P$ transitions are sufficient
in the low energy PNC scattering and also equally
important in both ways. Considering the lowest 
amplitudes, the derivative of the common factor 
in Eqs. \eqref{rotang} $\frac{d}{dz}\phi$ = 
Re$\mathcal{O}(k)$ and \eqref{pola} 
$\frac{d}{dz}P$ = Im$\mathcal{O}(k)$ 
(both per unit length) becomes 
\begin{equation}\label{obs}
\mathcal{O}(k)
=\frac{2i\pi M\mathcal{N}}{k^3}
\Bigl(
\mathcal{W}^{{}^3P_0}_{{}^1S_0}(k)
-\sqrt{2}\mathcal{W}^{{}^3P_1}_{{}^3S_1}(k)
-\mathcal{W}^{{}^1P_1}_{{}^3S_1}(k)
\Bigr),
\end{equation}
where the subscript ${}^3S_1$ includes also its
tensor coupled partner, the ${}^3D_1$ partial wave. 
The matrix elements of the PNC potential are further
written in terms of the matrix elements of the 
operators, which appear in the PNC potentials of the
DDH model and EFT, as    
\begin{equation}
\mathcal{W}^{{}^3P_0}_{{}^1S_0}(k)=
\frac{1}{M}\sum_\alpha\Bigl(
\mathcal{C}^{1\alpha}_{1[\times]_-}
\mathcal{J}^{{}^3P_0\alpha}_{{}^1S_0[\times]_-}(k)
+\mathcal{C}^{1\alpha}_{1[-]_+}
\mathcal{J}^{{}^3P_0\alpha}_{{}^1S_0[-]_+}(k)
\Bigr),
\end{equation}
\begin{equation}
\mathcal{W}^{{}^3P_1}_{{}^3S_1}(k)=
\frac{1}{M}\Bigl(\mathcal{C}^{1\pi}_{0[+]_-}
\mathcal{J}^{{}^3P_1\pi}_{{}^3S_1[+]_-}(k)
+\sum_\alpha\mathcal{C}^{1\alpha}_{0[+]_+}
\mathcal{J}^{{}^3P_1\alpha}_{{}^3S_1[+]_+}(k)\Bigr),
\end{equation}
\begin{equation}
\mathcal{W}^{{}^1P_1}_{{}^3S_1}(k)=
\frac{1}{M}\sum_\alpha\Bigl(
\mathcal{C}^{0\alpha}_{0[\times]_-}
\mathcal{J}^{{}^1P_1\alpha}_{{}^3S_1[\times]_-}(k)
+\mathcal{C}^{0\alpha}_{0[-]_+}
\mathcal{J}^{{}^1P_1\alpha}_{{}^3S_1[-]_+}(k)
\Bigr),
\end{equation}
where the meson label $\alpha$ $(=\rho,\omega)$ is for
the DDH model. In the case of EFT the $\alpha$'s and 
summation symbols are omitted. The reduced matrix
elements of the spin-space operators
\begin{align}\label{redmatj}
&\mathcal{J}^{{\mathcal{Q}'\alpha}}
_{\mathcal{Q}[\odot]_\pm}(k)
={}^{(-)}\langle k\hat{\bm{z}};\mathcal{Q}'||
(\bm{\sigma}_1\odot\bm{\sigma}_2)
\cdot[-i\bm{\nabla},Y_\alpha(r)]_\pm
||k\hat{\bm{z}};\mathcal{Q}\rangle^{(+)},
\end{align}
with ($\odot = \pm,\times$) are separated into the
commutator $[\odot]_-$ and anticommutator $[\odot]_+$
elements. The constants 
$\mathcal{C}^{T'\alpha}_{T[\odot]_\pm}$, where
$T$ denotes the total isospin in the initial and $T'$
in the final state, include the matrix elements of the
isospin operators and the other parameters associated
with the potential, {\it e.g.} in the case of the DDH
and pion, the constant is $\mathcal{C}^{1\pi}_{0[+]_-}
=g_\pi h_\pi^{(1)}/\sqrt{2}$.
\subsection{Neutron capture}
The thermal neutron capture cross-section on molecular
hydrogen does not depend on the interference or binding
effects of the protons in the molecule \cite{rie}. 
It is therefore sufficient to simply calculate the 
neutron capture cross-section on free protons.
The M1 ${}^1S_0\rightarrow{}^3S_1-{}^3D_1$ transition
dominates the $np\rightarrow\gamma d$ reaction at
threshold. By far the largest contribution 
(of about 90 \%) of this reaction arises from the 
impulse approximation which couples the $S$-states.
However, the one-pion exchange currents can also reach
the $D$-state of the deuteron and play an important 
role in explaining the experimental value of the 
cross-section for thermal neutrons as was shown in Ref.
\cite{riska}.

The relevant photoproduction vertices, in terms of
Lagrangian densities, are for the $\gamma NN\pi$
interaction
\begin{align}\label{photovertex}
\mathscr{L}_{\gamma NN\pi}&=
-e\frac{f_\pi}{m_\pi}\bar{N}
\gamma_5\gamma^\mu(\bm{\tau}
\times\bm{\pi})_zNA_\mu
\end{align}
and for the $\gamma\pi\pi$ interaction
\begin{equation}\label{piphotovertex}
\mathscr{L}_{\gamma\pi\pi}=
-e(\partial^\mu\bm{\pi}\times
\bm{\pi})_zA_\mu.
\end{equation}
\begin{figure} 
    \centering
\begin{tikzpicture}[scale=0.8]  
\draw (-4.8,-1.2) to [-] (-4.8,1.2); 
\draw (-3.2,-1.2) to [-] (-3.2,1.2);
\draw (-4.8,0) to [snake=coil,segment aspect=0,
line before snake=0pt,segment length=5pt] (-5.6,1.0);
\draw (-0.8,0) to [dashed] (0.8,0); 
\draw (-0.8,-1.2) to [-] (-0.8,1.2); 
\draw (0.8,-1.2) to [-] (0.8,1.2);
\draw (-0.8,0) to [snake=coil,segment aspect=0,
line before snake=0pt,segment length=5pt] (-1.6,1.0);
\draw (3.4,0) to [dashed] (5.0,0); 
\draw (5.0,-1.2) to [-] (5.0,1.2); 
\draw (3.4,-1.2) to [-] (3.4,1.2);
\draw (4.2,0) to [snake=coil,segment aspect=0,
line before snake=0pt,segment length=5pt] (4.2,1.2);
\draw (7.6,-0.5) to [dashed] (9.2,-0.5); 
\draw (7.6,-1.2) to [-] (7.6,1.2); 
\draw (9.2,-1.2) to [-] (9.2,1.2);
\draw (7.6,0.5) to [snake=coil,segment aspect=0,
line before snake=0pt,segment length=5pt] (7.0,1.2);
\draw (7.5,-0.5) to [-] (7.5,0.5);
\draw (7.6,-0.5) to [-] (7.5,-0.5);
\draw (7.6,0.5) to [-] (7.5,0.5);
\end{tikzpicture}
\caption{The diagrams for the magnetic dipole moments 
considered in the calculation of
$\sigma(np\rightarrow\gamma d)$. From left to right
they are called impulse (imp), seagull (sea), 
pion-in-flight (fly), and delta ($\Delta$) diagram. The
wavy line is a photon, the solid line is a nucleon, the
dashed line is a pion, and the bar is a 
$\Delta$-isobar.}
\label{npdgdiag}
\end{figure}
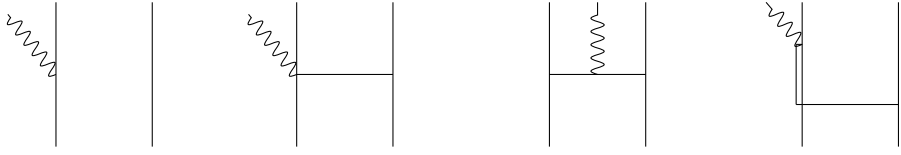
Since the energy of the resulting photon at threshold
of the reaction is only about 2 MeV, its wavelength is
much larger than the deuteron size, and thus the 
electric $\bm{E}$ and magnetic $\bm{B}$ fields can be
taken as constants. The scalar and vector potentials
of the uniform (static) fields $\bm{E}$ and
$\bm{B}$ are $\phi(\bm{r})=-\bm{E}\cdot\bm{r}$ and
$\bm{A}(\bm{r})=\frac{1}{2}\bm{B}\times\bm{r}$
respectively.

A diagrammatic illustration for the one- and two-body
magnetic dipole moment operators is given in Fig.
\ref{npdgdiag}. Besides Eqs. \eqref{photovertex} 
and \eqref{piphotovertex}, the other necessary $\gamma
N\Delta$, $\pi N\Delta$, and (PC and PNC) $\pi NN$
Lagrangians are given (in nonrelativistic form) in
Appendix \ref{apx}.
In the impulse approximation, the relevant spin changing
part of the operator is
\begin{equation}\label{magimp}
\hat{\bm{m}}^{\rm imp}=
\frac{\mu_v}{4}
(\hat{\tau}_{1z}-\hat{\tau}_{2z})
(\bm{\sigma}_1-\bm{\sigma}_2).
\end{equation}
The seagull contribution follows from Eqs. 
\eqref{photovertex} and \eqref{pcpinn} leading to 
the exchange operator
\begin{equation}\label{magsea}
\hat{\bm{m}}^{\rm sea}(\bm{r})
=-\frac{M}{2}\Bigl(\frac{f_\pi}{m_\pi}\Bigr)^2
(\bm{\tau}_1\times\bm{\tau}_2)_z
\Bigl\{
\hat{\bm{r}}[\hat{\bm{r}}\cdot
(\bm{\sigma}_1\times\bm{\sigma}_2)]
-\bm{\sigma}_1\times\bm{\sigma}_2
\Bigr\}(1+m_\pi r)
\frac{e^{-m_\pi r}}{4\pi r}.
\end{equation}
Similarly, the pion-in-flight contribution (the Sachs
exchange moment) is given by Eqs. \eqref{piphotovertex}
and \eqref{pcpinn}
\begin{align}\label{magfly}
&\hat{\bm{m}}^{\rm fly}(\bm{r})=
\nonumber\\
&-\frac{M}{2}
\Bigl(\frac{f_\pi}{m_\pi}\Bigr)^2
(\bm{\tau}_1\times\bm{\tau}_2)_z
\Bigl\{
\hat{\bm{r}}[\hat{\bm{r}}
\cdot(\bm{\sigma}_1\times
\bm{\sigma}_2)](1+m_\pi r)
+\bm{\sigma}_1\times
\bm{\sigma}_2(1-m_\pi r)
\Bigr\}\frac{e^{-m_\pi r}}{4\pi r}.
\end{align}
The final correction becomes by taking into account
the static $N\Delta$ intermediate state by using Eqs.
\eqref{pcpinn}, \eqref{gamndel}, and \eqref{pindel}
\begin{align}\label{magdel}
\hat{\bm{m}}^\Delta(\bm{r})=~&
\frac{\mu_\Delta f^\star_\pi f_\pi}
{9(M_\Delta-M)}
\Bigl((\hat{\tau}_{1z}-\hat{\tau}_{2z})
-i(\bm{\tau}_1\times\bm{\tau}_2)_z\Bigr) 
\Bigl\{i(\bm{\sigma}_1\times\hat{\bm{r}})
(\bm{\sigma}_2\cdot\hat{\bm{r}})
\nonumber\\
&-i(\bm{\sigma}_2\times\hat{\bm{r}})
(\bm{\sigma}_1\cdot\hat{\bm{r}})
-2(\bm{\sigma}_1-\bm{\sigma}_2)
\cdot\hat{\bm{r}}\hat{\bm{r}}
\Bigl\}
\Bigl(1+\frac{3}{m_\pi r}+
\frac{3}{(m_\pi r)^2}\Bigr)
\frac{e^{-m_\pi r}}{4\pi r},
\end{align}
where the irrelevant isospin conserving terms are 
omitted as in the case of $\hat{\bm{m}}^{\rm imp}$.
The total magnetic moment operator is given by
$\hat{\bm{m}}=\hat{\bm{m}}^{\rm imp}+
\hat{\bm{m}}^{\rm sea}+\hat{\bm{m}}^{\rm fly}+
\hat{\bm{m}}^\Delta$.

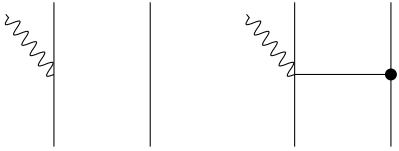
\begin{figure} 
    \centering
\begin{tikzpicture}[scale=0.8]  
\draw (-4.8,-1.2) to [-] (-4.8,1.2); 
\draw (-3.2,-1.2) to [-] (-3.2,1.2);
\draw (-4.8,0) to [snake=coil,segment aspect=0,
line before snake=0pt,segment length=5pt] (-5.6,1.0);
\node (b) at (0.8,0) 
[fill,circle,inner sep=1.6pt] {};
\draw (-0.8,0) to [dashed] (0.8,0); 
\draw (-0.8,-1.2) to [-] (-0.8,1.2); 
\draw (0.8,-1.2) to [-] (0.8,1.2);
\draw (-0.8,0) to [snake=coil,segment aspect=0,
line before snake=0pt,segment length=5pt] (-1.6,1.0);
\end{tikzpicture}
\caption{The E1 contributions to the $\gamma$-asymmetry 
$\mathcal{A}_\gamma$. The black dot denotes the parity
nonconserving vertex.}
\label{e1diagrams}
\end{figure}

The nonzero $\gamma$-asymmetry $\mathcal{A}_\gamma$
arises from the interference between the M1 transition
and the PNC interaction propelled E1 transitions. 
The E1 transitions connect the initial ${}^3S_1-{}^3D_1$
and final deuteron states through the parity admixed
continuum and bound $\widetilde{{}^3P_1}$ states, while
the PNC E1 transitions connect them directly. 
Figure \ref{e1diagrams} gives the diagrams for 
electric dipole moment contributions. The one-body
operator is given by 
\begin{equation}
\hat{\bm{\mu}}^{\rm imp}_e=
\frac{e}{4}
(\hat{\tau}_{1z}-\hat{\tau}_{2z})
\bm{r}
\end{equation}
and the two-body PNC exchange operator results from
Eqs. \eqref{photovertex} and \eqref{pncnnpi}
\begin{equation}\label{pnce1}
\hat{\bm{\mu}}^{\rm PNC}_e(\bm{r})
=\frac{e}{4M}
\frac{f_\pi h_\pi^{(1)}}{m_\pi\sqrt{2}}
(\bm{\tau}_1\cdot\bm{\tau}_2
-\hat{\tau}_{1z}\hat{\tau}_{2z})
\Bigl[i(\bm{\sigma}_1+\bm{\sigma}_2)
\frac{e^{-m_\pi r}}{4\pi r}
-\bm{r}(\bm{\sigma}_1+\bm{\sigma}_2)\cdot
\Bigl\{-i\bm{\nabla},
\frac{e^{-m_\pi r}}{4\pi r}\Bigr\}
\Bigr].
\end{equation}
The total E1 operator is the sum
$\hat{\bm{\mu}}_e =
\hat{\bm{\mu}}^{\rm imp}_e + 
\hat{\bm{\mu}}^{\rm PNC}_e$.
Note that the $\hat{\bm{\mu}}^{\rm imp}_e$ changes
parity, while the $\hat{\bm{\mu}}^{\rm PNC}_e$
conserves it. 
Reference \cite{simonius} investigates the PNC E1
transitions and gives an additional $\gamma NN\pi$
vertex leading to the spin-changing PNC E1 operator.
However, this vertex has a vanishing contribution to 
the $\mathcal{A}_\gamma$ because the term proportional
to the $\bm{B}$ is negligibly small and the M1-E1
interference disappears as a consequence of the 
initial ${}^1S_0$ states in both amplitudes. 

In terms of the reduced matrix elements of the electric
and magnetic dipole transition operators, the
$\gamma$-asymmetry $\mathcal{A}_\gamma=
(d\sigma_{+\frac{1}{2}}-d\sigma_{-\frac{1}{2}})/
(d\sigma_{+\frac{1}{2}}+d\sigma_{-\frac{1}{2}})$ reads
now
\begin{align}\label{gamasym}
\mathcal{A}_\gamma(k)=
\sqrt{2}{\rm Re}\Bigl[\frac{i\sum_{LL^d}
\int_0^\infty dr\mathcal{D}_{L^d}(r)
\langle{}^3L_1^d||
\bm{\hat{\mu}}_e(\bm{r})||
{}^3L_1\rangle
\mathcal{U}_{{}^3L_1}^{(+)}(r,k)}
{\sum_{L^d}
\int_0^\infty dr\mathcal{D}_{L^d}(r)
\langle{}^3L_1^d||
\hat{\bm{\mu}}_m(\bm{r})||
{}^1S_0\rangle
\mathcal{U}_{{}^1S_0}^{(+)}(r,k)
}\Bigr],
\end{align}
where $\hat{\bm{\mu}}_m =e\hat{\bm{m}}/2M$ and the
multiplication factor $\cos\theta$, in which
$\theta$ is the angle between the neutron spin and
photon direction, is left out. 
\section{\label{res}Results}
The results for the neutron spin rotation 
$\frac{d}{dz}\phi$, polarization $\frac{d}{dz}P$, and
$\gamma$-asymmetry $\mathcal{A}_\gamma$ are given in
this section. The $NN$ wavefunctions are the exact
solutions of the coupled Schr\"{o}dinger equations
where the strong distortions originate from the Reid93
potential \cite{reid93}. The OME parts of the 
observables can be and are, as customary, evaluated
without regularization.
The $\frac{d}{dz}\phi$ and $\frac{d}{dz}P$ are low
energy scattering problems which are handled
perturbatively within the DWBA and therefore the
inclusion of the form factors is not crucial. The DWBA
approach allows the PNC potentials to be omitted from
the Schr\"{o}dinger equations because of their
diminutive distortive effects to the strong
wavefunctions.
This, however, is not possible in the calculation of
the $\mathcal{A}_\gamma$ since the parity admixed
wavefunctions are needed. 
In this case also the regularization of the PNC TPE
potentials becomes necessity for the reason that they
are too singular to be treated without it. 
As for the DDH model results, the DDH "best" values
for the weak couplings are used along with the strong
couplings $g_\pi=13.45$, $g_\rho=2.79$, and 
$g_\omega=8.37$ and the anomalies $\chi_\rho=3.71$
and $\chi_\omega=-0.12$. 

The employed TPE potentials are taken from Refs.
\cite{k2pi} and \cite{d2pi} and respectively 
abbreviated as $\hat{V}_{\rm K}^{\rm TPE}(\bm{r})$
and $\hat{V}_{\rm D}^{\rm TPE}(\bm{r})$ of which the
former takes into account the $NN$ and $N\Delta$
intermediate states while the latter only the $NN$.
The $NN$ intermediate parts of these potentials
coincide, if the constant term in the dispersion 
relation of the $\hat{V}_{\rm D}^{\rm TPE}(\bm{r})$
is excluded. The constant term in momentum space 
corresponds to a term proportional to the delta
function $\delta(\bm{r})$ in coordinate space.
However, when the TPE potentials are regularized by
form factors, their $NN$ parts inevitably differ
from each other.
The TPE potentials are modified by the monopole 
$\Lambda^2(\bm{q}^2+\Lambda^2)^{-1}$ and dipole
$\Lambda^4(\bm{q}^2+\Lambda^2)^{-2}$ form factors and
used with three different cut-off mass values 
$\Lambda=$ 0.8, 1.0, and 1.2 GeV. The same type of
the monopole form factor emerged with the PNC TPE
potential $\hat{V}_{\rm D}^{\rm TPE}(\bm{r})$
was also used in Ref. \cite{hyunncap} to calculate the 
$\gamma$-asymmetry in $\vec{n}p\rightarrow\gamma d$
at threshold. The results of the medium-range TPE are 
sensitive to the form factors, which have an
increasingly suppressing effect on them when the 
cut-off is decreased and the rank of the form factor
is raised. 
\subsection{Neutron spin rotation $\frac{d}{dz}\phi$
and polarization $\frac{d}{dz}P$}  
The spin-space operators appearing in Eq. 
\eqref{redmatj} are common building blocks in the
PNC $NN$ interaction. In the approach in which the
operators are placed between high-quality 
phenomenological wavefunctions, the low-energy 
$\lesssim$ 1 keV scattering matrix elements can be
expressed in constant form, as given in Tab. 
\ref{matelms}. Therefore, by means of the matrix 
elements of Tab. \ref{matelms}, it is 
straightforward to customize the OME contributions
within any model that uses the $\pi$, $\omega$,
and/or $\rho$ exchanges as the ranges of the 
unmodified Yukawa functions
$Y_\alpha(r)=e^{-m_\alpha r}/4\pi r$.
\begin{table}[b]
\caption{\label{matelms} 
The values of Eq. \eqref{redmatj} 
for the spin rotation components
$\mathcal{I}^{\mathcal{Q}'\alpha}
_{\mathcal{Q}[\times]_-}(k)={\rm Re}\Bigl(k^{-3}
\mathcal{J}^{\mathcal{Q}'\alpha}
_{\mathcal{Q}[\times]_-}(k)\Bigr)$ and 
$\mathcal{I}^{\mathcal{Q}'\alpha}
_{\mathcal{Q}[\pm]_\pm}(k)={\rm Re}\Bigl(-ik^{-3}
\mathcal{J}^{\mathcal{Q}'\alpha}
_{\mathcal{Q}[\pm]_\pm}(k)\Bigr)$
in units of mb and for the spin polarization components
$\mathcal{G}^{\mathcal{Q}'\alpha}
_{\mathcal{Q}[\times]_-}(k)=\frac{k}{T_{\rm Lab}^n}
{\rm Im}\Bigl(k^{-3}\mathcal{J}^{\mathcal{Q}'\alpha}
_{\mathcal{Q}[\times]_-}(k)\Bigr)$ and
$\mathcal{G}^{\mathcal{Q}'\alpha}
_{\mathcal{Q}[\pm]_\pm}(k)=\frac{k}{T_{\rm Lab}^n}
{\rm Im}\Bigl(-ik^{-3}\mathcal{J}^{\mathcal{Q}'\alpha}
_{\mathcal{Q}[\pm]_\pm}(k)\Bigr)$ in units
of $10^{-10}~{\rm fm}/{\rm meV}$. 
The $k$ and $T_{\rm Lab}^n$ (related as
$T_{\rm Lab}^n=\frac{2k^2}{M}$) are respectively in
units of $\rm{fm}^{-1}$ and meV. The functions
$\mathcal{I}$ and $\mathcal{G}$ are constants within
the neutron kinetic energy range of about 
$T_{{\rm Lab}}^n$=0-1 keV.}
\begin{tabular}{|c||c|c|c|}
\hline\hline
\cline{2-4}
&$\pi$
&$\rho$
&$\omega$\\
\hline\hline
$\mathcal{I}^{{}^3P_0\alpha}
_{{}^1S_0[\times]_-}$
&~$-30.1992$~
&~$-0.7825$~
&~$-0.7444$~\\
$\mathcal{I}^{{}^3P_0\alpha}
_{{}^1S_0[-]_+}$
&~$-42.2133$~
&~$-0.6494$~
&~$-0.6143$~\\
$\mathcal{I}^{{}^1P_1\alpha}
_{{}^3S_1[\times]_-}$
&~$1.5588$~
&~$0.0868$~
&~$0.0834$~\\
$\mathcal{I}^{{}^1P_1\alpha}
_{{}^3S_1[-]_+}$
&~$-4.4409$~
&~$-0.1020$~
&~$-0.0975$~\\
$\mathcal{I}^{{}^3P_1\alpha}
_{{}^3S_1[+]_-}$
&~$3.4753$~
&~$0.1528$~
&~$0.1459$~\\
$\mathcal{I}^{{}^3P_1\alpha}
_{{}^3S_1[+]_+}$
&~$5.6818$~
&~$0.0843$~
&~$0.0798$~\\
\hline\hline
\end{tabular}
\quad
\begin{tabular}{|c||c|c|c|}
\hline\hline
\cline{2-4}
&$\pi$
&$\rho$
&$\omega$\\
\hline\hline
$\mathcal{G}^{{}^3P_0\alpha}
_{{}^1S_0[\times]_-}$
&~$-8.6220$~
&~$-0.2234$~
&~$-0.2125$~\\
$\mathcal{G}^{{}^3P_0\alpha}
_{{}^1S_0[-]_+}$
&~$-12.0522$~
&~$-0.1854$~
&~$-0.1754$~\\
$\mathcal{G}^{{}^1P_1\alpha}
_{{}^3S_1[\times]_-}$
&~$-0.1020$~
&~$-0.0057$~
&~$-0.0055$~\\
$\mathcal{G}^{{}^1P_1\alpha}
_{{}^3S_1[-]_+}$
&~$0.2905$~
&~$0.0067$~
&~$0.0064$~\\
$\mathcal{G}^{{}^3P_1\alpha}
_{{}^3S_1[+]_-}$
&~$-0.2273$~
&~$-0.0100$~
&~$-0.0095$~\\
$\mathcal{G}^{{}^3P_1\alpha}
_{{}^3S_1[+]_+}$
&~$-0.3717$~
&~$-0.0055$~
&~$-0.0052$~\\
\hline\hline
\end{tabular}
\end{table}

The neutron spin rotation is split into one-meson
exchange (OME) and two-pion exchange (TPE) components
as $\frac{d}{dz}\phi=\frac{d}{dz}\phi^{\rm OME}
+\frac{d}{dz}\phi^{\rm TPE}$. 
Since a neutron scatters coherently from a pair of
protons in the hydrogen molecule, it is then 
appropriate to use the molecule number density of
liquid hydrogen instead of a two times larger 
atom number density. Therefore, the liquid 
parahydrogen particle density value of 
$\mathcal{N}=0.021$ molecules/$\mathring{A}^3$ is
used in the numerical results. In terms of the DDH
model, the rotation may be written as 
\begin{equation}\label{ddhrot}
\frac{d}{dz}\phi^{\rm OME}_{\rm DDH}
=\Bigl(0.617h^{(1)}_\pi
-0.138h^{(0)}_\omega
-0.012h^{(1)}_\omega
-0.126h^{(0)}_\rho
+0.004h^{(1)}_\rho
+0.130h^{(2)}_\rho\Bigr)\frac{\rm rad}{\rm m}.
\end{equation}
This has the value of
$3.31\times10^{-7}$ $\frac{\rm rad}{\rm m}$ when the
DDH "best" values are plugged in. With this model and
values, the one-pion exchange (OPE) has a dominance of
about 85\%. 
The result of Eq. \eqref{ddhrot} is consistent
with the one in Ref. \cite{liutimm}, except half the
size (because of the half the size particle density
value), and also in line with the predictions of Refs.
\cite{schia,liu07} which all employ the Argonne 
$v_{18}$ potential.
The result using the Paris potential reported in Ref.
\cite{avis2} is in the same order with the 
aforementioned results but of the opposite sign.
Also in the result of Ref. \cite{avis2} some of the
signs between the partial contributions are in
disagreement with the mutually consistent results of
Ref. \cite{liutimm} and Eq. \eqref{ddhrot}. 

\begin{table}[ht]
\caption{\label{twopirot} 
The TPE contributions to the neutron spin rotation
$\frac{d}{dz}\phi^{\rm TPE}$ in units of
$h^{(1)}_\pi\frac{\rm rad}{\rm m}$ and polarization 
$\frac{k}{T_{\rm Lab}^n}\frac{d}{dz}P^{\rm TPE}$ 
in units of $h^{(1)}_\pi\times
10^{-12}~{\rm fm}^{-1}{\rm m}^{-1}{\rm meV}^{-1}$.
The K and D stand for the
$\hat{V}_{\rm K}^{\rm TPE}(\bm{r})$ and 
$\hat{V}_{\rm D}^{\rm TPE}(\bm{r})$ while the F and
FF stand respectively for the modification by
$\Lambda^2(\bm{q}^2+\Lambda^2)^{-1}$ and 
$\Lambda^4(\bm{q}^2+\Lambda^2)^{-2}$, where the 
cut-off masses $\Lambda$ are in units of GeV.}
\begin{tabular}{|c||c|c|c|c|}
\hline\hline
\multicolumn{5}{|c|}{$\frac{d}{dz}\phi^{\rm TPE}$} \\
\hline
\cline{2-5}
$\Lambda$
&K(FF)
&D(FF)
&K(F)
&D(F)\\
\hline\hline
$0.8$
&~$-0.040$~
&~$0.026$~
&~$-0.105$~
&~$-0.020$~\\
$1.0$
&~$-0.074$~
&~$0.004$~
&~$-0.132$~
&~$-0.040$~\\
$1.2$
&~$-0.103$~
&~$-0.017$~
&~$-0.152$~
&~$-0.055$~\\
\hline\hline
\end{tabular}
\quad
\begin{tabular}{|c||c|c|c|c|}
\hline\hline
\multicolumn{5}{|c|}{
$\frac{k}{T_{\rm Lab}^n}\frac{d}{dz}P^{\rm TPE}$} \\
\hline
\cline{2-5}
$\Lambda$
&K(FF)
&D(FF)
&K(F)
&D(F)\\
\hline\hline
$0.8$
&~$2.648$~
&~$-1.712$~
&~$6.845$~
&~$1.295$~\\
$1.0$
&~$4.844$~
&~$-2.442$~
&~$8.648$~
&~$2.619$~\\
$1.2$
&~$6.763$~
&~$1.104$~
&~$9.925$~
&~$3.604$~\\
\hline\hline
\end{tabular}
\end{table}

Because the $\hat{V}_{\rm D}^{\rm TPE}(\bm{r})$ and $NN$
part of the $\hat{V}_{\rm K}^{\rm TPE}(\bm{r})$
are identical in unregularized form, they contribute
the same amount to the rotation 
$\frac{d}{dz}\phi^{\rm TPE}_{NN}=-0.101h^{(1)}_\pi$
$\frac{\rm rad}{\rm m}$ reducing the OPE effect 
by about $15\%$. By taking into account also the
$\Delta$ effects of the 
$\hat{V}_{\rm K}^{\rm TPE}(\bm{r})$, the contribution
doubles to
$\frac{d}{dz}\phi^{\rm TPE}_{\rm K}=-0.209h^{(1)}_\pi$
$\frac{\rm rad}{\rm m}$ cutting down the OPE effect by 
over $30\%$.  
However, since the TPE is a medium-range effect, it is
sensitive to form factors. The form factor modified TPE
contributions to the $\frac{d}{dz}\phi^{\rm TPE}$ are
given in Tab. \ref{twopirot}. As is seen in the Table,
the regularized TPE contributions are, like the 
unregularized ones, negative in almost all cases and,
thus, opposite to the OPE effect. Only the D(FF) case
differs by having a different sign at small cut-off
masses.

When it comes to the neutron spin polarization
$\frac{d}{dz}P$, the effect is much smaller than the
one of $\frac{d}{dz}\phi$ and therefore not 
particularly interesting because it is experimentally
much less achievable. For example, similarly to
Eq. \eqref{ddhrot}, the DDH OME contribution to the 
polarization with $T_{\rm Lab}^n$ = 10 meV neutrons is
only $\frac{d}{dz}P^{\rm OME}_{\rm DDH}=
7.43\times10^{-12}$ $\frac{1}{m}$.
This result can be constructed by using the matrix
elements $\mathcal{G}^{\mathcal{Q}'\alpha}
_{\mathcal{Q}[\odot]_\pm}(k)$ of Tab. \ref{matelms},
which lead with the DDH "best" values to
$\frac{k}{T_{\rm Lab}^n}
\frac{d}{dz}P^{\rm OME}_{\rm DDH}=8.17\times
10^{-18}~{\rm fm}^{-1}{\rm m}^{-1}{\rm meV}^{-1}$.
The OPE contribution to this number in the same units
is $-18.54$ and has, thus, about $70\%$ dominance. 
As in the rotation, the unregularized TPE contributions
to the $\frac{k}{T_{\rm Lab}^n}\frac{d}{dz}P^{\rm TPE}$
(in units of $h^{(1)}_\pi\times
10^{-12}~{\rm fm}^{-1}{\rm m}^{-1}{\rm meV}^{-1}$) are
6.63 and 13.70, in which the former is the same for
both potentials when only the $NN$ intermediate states
are considered and the latter is for the
$\hat{V}_{\rm K}^{\rm TPE}(\bm{r})$ when also the
$N\Delta$ intermediate states are included. As in the
corresponding case of the rotation, the TPE effect with
the $\Delta$, diminish over $30\%$ the OPE contribution.
The effects of the form factors on the
$\frac{k}{T_{\rm Lab}^n}\frac{d}{dz}P^{\rm TPE}$
are presented in Tab. \ref{twopirot} and show similar
features as in the case of the spin rotation. 
\subsection{Photon asymmetry $\mathcal{A}_\gamma$}
An experimental value of the radiative thermal neutron
capture cross-section of proton is 
$\sigma(np\rightarrow\gamma d)=334.2 \pm 0.5$ mb 
\cite{cox}. Theoretically it may be given by
\begin{align}\label{abscs}
\sigma_{\rm cap}=
\frac{\alpha\pi\omega_\gamma^3}{6k^3M}&
\sum_{L_d}\Big|
\int_0^\infty dr\mathcal{D}_{L_d}(r)
\langle{}^3L_{d1}||
\hat{\bm{m}}(\bm{r})
||{}^1S_0\rangle
\mathcal{U}_{{}^1S_0}^{(+)}(r,k)\Big|^2,
\end{align}
where $\alpha$ $(=e^2)$ is the fine structure constant,
$\omega_\gamma$ the C.M. energy of the photon, and
$\hat{\bm{m}}(\bm{r})$ the magnetic moment operator,
which is the sum of Eqs. \eqref{magimp}-\eqref{magdel}. 
The used coupling values are 
$g_\pi=13.45~(f_\pi=m_\pi g_\pi/2M)$,
$f^\star_\pi=\sqrt{72/25}f_\pi$, and
$\mu_\Delta=f^\star_\pi\mu_v/2f_\pi$
$(f_{\gamma N\Delta}/m_\pi=e\mu_\Delta/2M)$, 
where $g_\pi$ is the $\pi NN$ coupling, 
$f^\star_\pi$ the quark model result for the
$\pi N\Delta$ coupling \cite{npidel}, 
$\mu_\Delta$ the transition magnetic moment, and 
$\mu_v=4.71$ the isovector magnetic moment of the
nucleon. 
Evaluation of Eq. \eqref{abscs} for the thermal 
neutrons ($T_{{\rm Lab}}^n =25$ meV) gives 
$\sigma_{\rm cap}=334.42$ mb, which is in an
excellent agreement with the experimental 
cross-section. In this result, the mere impulse
approximation, {\it i.e.}
$\hat{\bm{m}}(\bm{r})=\hat{\bm{m}}^{\rm imp}$, 
produces the largest part of the cross-section 
giving $\sigma_{\rm cap}^{\rm imp}$ = 303.81 mb.
The missing $\sim$ 10\% enhancement results from
the OPE current corrections, as proposed in Ref.
\cite{riska}. 

By using the DDH model and their "best" coupling
values, the OME contribution to the $\gamma$-asymmetry 
Eq. \eqref{gamasym} is
$\mathcal{A}_{\gamma \rm DDH}^{\rm OME} 
=-5.387\times 10^{-8}$.
Even though included, the effects of the PNC E1 current
($\sim -0.2\permil$) and heavy meson $\rho$- and $\omega$-exchanges (less than $1\%$) are negligibly
small. In terms of the weak pion coupling, the result
may be written as 
$\mathcal{A}_{\gamma \rm DDH}^{\rm OME} 
\approx-0.117h^{(1)}_\pi$ which is in harmony with the
previous predictions (see the references mentioned in
the Introduction) in most cases. The results, in which
the TPE contributions are added on top of the 
$\mathcal{A}_{\gamma \rm DDH}^{\rm OME} $, are shown
in Tab. \ref{twopiabs} and are rather self-explaining. 
The result of the model D(F) with $\Lambda=1.0$ GeV is consistent with Ref. \cite{hyunncap} and show $\sim6\%$
smaller $\gamma$-asymmetry than without the TPE.
Just like in scattering cases, the TPE contribution 
of the D(FF) model with $\Lambda=0.8$ and 1.0 GeV
differs by the sign from the other models and,
therefore, strengthens the total asymmetry. 
Otherwise the TPE effect is destructive as in the
case of the $\frac{d}{dz}\phi^{\rm TPE}$. 
The largest contribution arises from the K(F) model
with $\Lambda=1.2$ GeV, which diminishes the
$\mathcal{A}_{\gamma \rm DDH}^{\rm OME}$ of about
about 20\%.

\begin{table}[ht]
\caption{\label{twopiabs} 
Photon asymmetry $\mathcal{A}_\gamma$ 
including the TPE in units of $10^{-8}$.}
\begin{tabular}{|c||c|c|c|c|}
\hline\hline
\multicolumn{5}{|c|}{
$\mathcal{A}_\gamma(\vec{n}p\rightarrow\gamma d)$} \\
\hline
\cline{2-5}
$\Lambda$
&K(FF)
&D(FF)
&K(F)
&D(F)\\
\hline\hline
$0.8$
&~$-5.08$~
&~$-5.58$~
&~$-4.60$~
&~$-5.24$~\\
$1.0$
&~$-4.83$~
&~$-5.41$~
&~$-4.40$~
&~$-5.08$~\\
$1.2$
&~$-4.61$~
&~$-5.26$~
&~$-4.26$~
&~$-4.97$~\\
\hline\hline
\end{tabular}
\end{table}
\section{Summary}\label{summar}
Three PNC observables for cold neutron interaction with
parahydrogen were calculated. All the observables, the
neutron spin rotation $\frac{d}{dz}\phi$ and 
polarization $\frac{d}{dz}P$ in scattering, and the
$\gamma$-asymmetry $\mathcal{A}_\gamma$ in capture,
were found to be dominated by the pion exchange. The
effect of the TPE was also taken into account in these 
observables and investigated in several settings. 
In all cases, the TPE effect was mainly opposite to
the one of the OPE. 

The OME contribution to the $\frac{d}{dz}\phi$ was
concluded to be a factor of two smaller than the 
most recent predictions and that the TPE decreased it
up to $\sim$ 30\% further. The $\frac{d}{dz}P$ was
considered rather uninteresting due to its small size.
In the $\mathcal{A}_\gamma$,  
the OPE currents gave the expected increment for the
M1 transitions but were insignificant for PNC E1
ones. The asymmetry was found to arise more or
less completely from the pion exchange, {\it i.e.} the
OPE weakened by the TPE up to $\sim$ 20\% or so.

Unfortunately, so far there exist no direct and 
helpful experimental data for these specific 
observables. However, as already mentioned, there exist
three precision experiment data points of the PNC
longitudinal analyzing power
$\bar{A}_L(\vec{p}p\rightarrow pp)$ of which two are
low energy points originating completely from the 
${}^1S_0-{}^3P_0$ transition. A recent $\bar{A}_L$
calculation of Ref. \cite{pp2} uses exactly the same
TPE models (potentials, couplings, and form factors)
as used here. 
Based on the DDH model along with the TPE model K(FF)
with $\Lambda=1.0$ GeV gives the best fit and is also
in a good agreement with the experimental data. This
suggests that, in the calculated observables 
$\frac{d}{dz}\phi$, $\frac{d}{dz}P$, and
$\mathcal{A}_\gamma$, the destructive TPE effect
compared to the OPE one is roughly $10\%$ in each 
case. 
\begin{acknowledgements}
The author is grateful to Dr. J. A. Niskanen
for advice and reading the manuscript.
The financial support from Vilho, Yrj\"{o} ja Kalle
V\"{a}is\"{a}l\"{a} Foundation is also gratefully
acknowledged. 
\end{acknowledgements}
\appendix
\section{}\label{apx}
The nonrelativistic interaction Lagrangians 
(the couplings are explained in the text) are
\begin{equation}\label{pcpinn}
L_{\pi NN}=
\frac{f_\pi}{m_\pi}
\bm{\sigma}\cdot\bm{\nabla}
\bm{\tau}\cdot\bm{\pi},
\end{equation}
\begin{equation}\label{gamndel}
L_{\gamma N\Delta}=
\frac{f_{\gamma N\Delta}}{m_\pi}
\bm{S}\cdot\bm{B}\hat{T}_z+{\rm H.c.},
\end{equation}
\begin{equation}\label{pindel}
L_{\pi N\Delta}=
\frac{f^\star_\pi}{m_\pi}
\bm{S}\cdot\bm{\nabla}\bm{T}
\cdot\bm{\pi}+{\rm H.c.},
\end{equation}
\begin{equation}\label{pncnnpi}
L_{\pi NN}^{\rm PNC}=
\frac{h_\pi^{(1)}}{\sqrt{2}}
(\bm{\tau}\times\bm{\pi})_z.
\end{equation}
The $\bm{S}$ and $\bm{T}$ are respectively the 
$N\Delta$ spin and isospin transition operators
\cite{npidel}.
\newpage
%
%
%
%
\end{document}